\documentclass[pra,twocolumn,10pt,aps]{revtex4-1}
\usepackage[dvips]{graphics,graphicx}
\usepackage{amsmath}

\usepackage{amssymb}
\usepackage[usenames, dvipsnames]{color}
\usepackage{graphicx}
\usepackage{graphics}
\usepackage{hyperref}
\usepackage[table, dvipsnames]{xcolor}

\usepackage{amsfonts}
\usepackage{eufrak}
\usepackage{mathrsfs}
\usepackage{bm}

\usepackage{soul}
\usepackage{soulutf8}

\newcommand{\scs}{\scriptscriptstyle}

\definecolor{1color}{RGB}{255,170,204}

\definecolor{2color}{RGB}{42,212,255}

\definecolor{3color}{RGB}{141,211,95}

\definecolor{4color}{RGB}{255,179,128}

\begin{document}

\title{Electrically controlled laser generation \\ 
in a photonic crystal - liquid crystal - metal microcavity}
\author{
Daniil~S.~Buzin$^{1,\dagger}$, Pavel~S.~Pankin$^{1,2,\dagger,*}$, Dmitrii~N.~Maksimov$^{1,2,\dagger}$,
Vitaly~S.~Sutormin$^{1,2,\dagger}$, Gavriil~A.~Romanenko$^{3}$,
Rashid~G.~Bikbaev$^{1,2}$,
Sergey~V.~Nedelin$^{1,4}$, Nikita~A.~Zolotovskii$^{1,4}$,
Igor~A.~Tambasov$^{1,4}$, Stepan~Ya.~Vetrov$^{2,1}$, Kuo-Ping~Chen$^{5,6}$, 
and Ivan~V.~Timofeev$^{1,2}$}
\affiliation{$^{1}$Kirensky Institute of Physics, Krasnoyarsk Scientific Center, Siberian Branch, Russian Academy of Sciences, Krasnoyarsk, 660036 Russia}
\affiliation{$^{2}$Siberian Federal University, Krasnoyarsk, 660041 Russia}
\affiliation{$^{3}$Faculty of Physics of ITMO University, St. Petersburg, 197101, Russia}
\affiliation{$^{4}$LLC Research and Production Company “Spectehnauka”, 660043 Krasnoyarsk, Russia}
\affiliation{$^{5}$Institute of Photonics Technologies, National Tsing Hua University, Hsinchu, Taiwan}
\affiliation{$^{6}$College of Photonics, National Yang Ming Chiao Tung University, 301 Gaofa 3rd Road, Tainan 711, Taiwan}
\affiliation{$^{\dagger}$Contributed equally to this work}
\affiliation{$^{*}$Corresponding author: pavel-s-pankin@iph.krasn.ru}
\date{\today}

\begin{abstract}
A comprehensive approach
for simulating lasing dynamics in a liquid crystal based laser is presented.
The approach takes into account the transformation of the liquid crystal structure caused by applied voltage. In particular, it allows us to explicitly account for a resonant mode frequency shift in the laser equations.
The laser dynamic is described by a set of coupled non-linear differential equations for dye polarizations, population densities and the electromagnetic fields. 
The proposed model is applied to a photonic crystal$-$metal microcavity filled with a resonant nematic liquid crystal layer doped with a dye.
The calculated lasing spectra governed by external electric field are verified in comparison with measured spectra. 
\end{abstract}


\maketitle



\section{INTRODUCTION}

The tunability is a key feature of the liquid crystal (LC) lasers that can be achieved due to  sensitivity of LCs to external factors~\cite{mysliwiec2021liquid}.
Tunable lasers based on the photonic crystal (PhC) microcavity with a LC resonant layer doped with dye molecules were firstly demonstrated in~\cite{ozaki2003electrically, ozaki2004electrically}.
Later, it was shown that the wavelength and polarization of light emitted from a vertical cavity surface emitting laser (VCSEL) based on PhC and PhC/metal microcavities can be controlled by using an embedded LC layer~\cite{castany2011tunable, panajotov2011vertical, xie2012vcsel, frasunkiewicz2018electrically, boisnard2020cw}.
The PhC/LC/PhC~\cite{krasnov2023voltage} and PhC/LC/metal~\cite{pankin2020one, romanenko2023metal} microcavities are known to be a versatile platform for engineering resonances with tunable Q-factors induced by bound states in the continuum.
Lasers based on bound states in the continuum have low-threshold~\cite{yang2021low} and provide non-trivial topology for emitted light~\cite{wang2024optical}. 
Lasers with self-organized photonic band gap structures, such as cholesteric LCs~\cite{kopp1998low, coles2010liquid, huang2016electrically, jeong2024quickly}, are a perspective platform for active beam steering \cite{zhang2023non, cho2020emission}.

The lasing dynamics in photonic systems is usually simulated by the finite difference time domain method (FDTD)~\cite{chang2004finite}. 
The response of the dopant molecules in dye lasers can be described using a four level system approximation~\cite{azzam2019exploring, arnold2013modeling, prokopevacloud}.
In this approach the transitions between the energy levels are governed by
coupled rate equations with phenomenological radiation lifetimes.
The material polarization induced by dipole transitions  is taken into account in Maxwell's equations for the elecromagnetic (EM) field in the microcavity. 
The same model was applied to a cholesteric LC for simulating lasing from edge~\cite{matsui2010finite} and defect~\cite{matsui2012finite} modes.
However, the sensitivity of LC to external factors was taken into account by phenomenological changing a cholesteric LC pitch without full-fledged modeling of the LC structure transformation under external factors.

In this work we present a comprehensive approach for simulating lasing dynamics in LC based systems.
In the first step the transformation of the LC structure under an applied voltage is calculated in the framework of the Frank-Oseen model~\cite{blinov2010structure} with the finite LC anchoring energy described by the Rapini potential~\cite{muravsky2024q}.
In the second step the corresponding change in the LC permittivity is taken into account in Maxwell's equations, which are solved  consistently with the equations for the emitters' level population and polarizations.
The equations for emitters' energy level population dynamics are derived from the Markovian master equation in the Lindblad form~\cite{scully1997quantum}.
In contrast to the earlier works~\cite{matsui2010finite,matsui2012finite} our approach does not contain phenomenological radiation decay lifetimes.
The final set of equations is solved numerically by using FDTD method on the Yee grid.
The theoretical results are compared with the experimentally measured electrically controlled lasing spectra of a PhC/LC/metal microlaser.

\section{THE MODEL}

\begin{figure}[t]
    \centering    \includegraphics[width=0.9\linewidth]{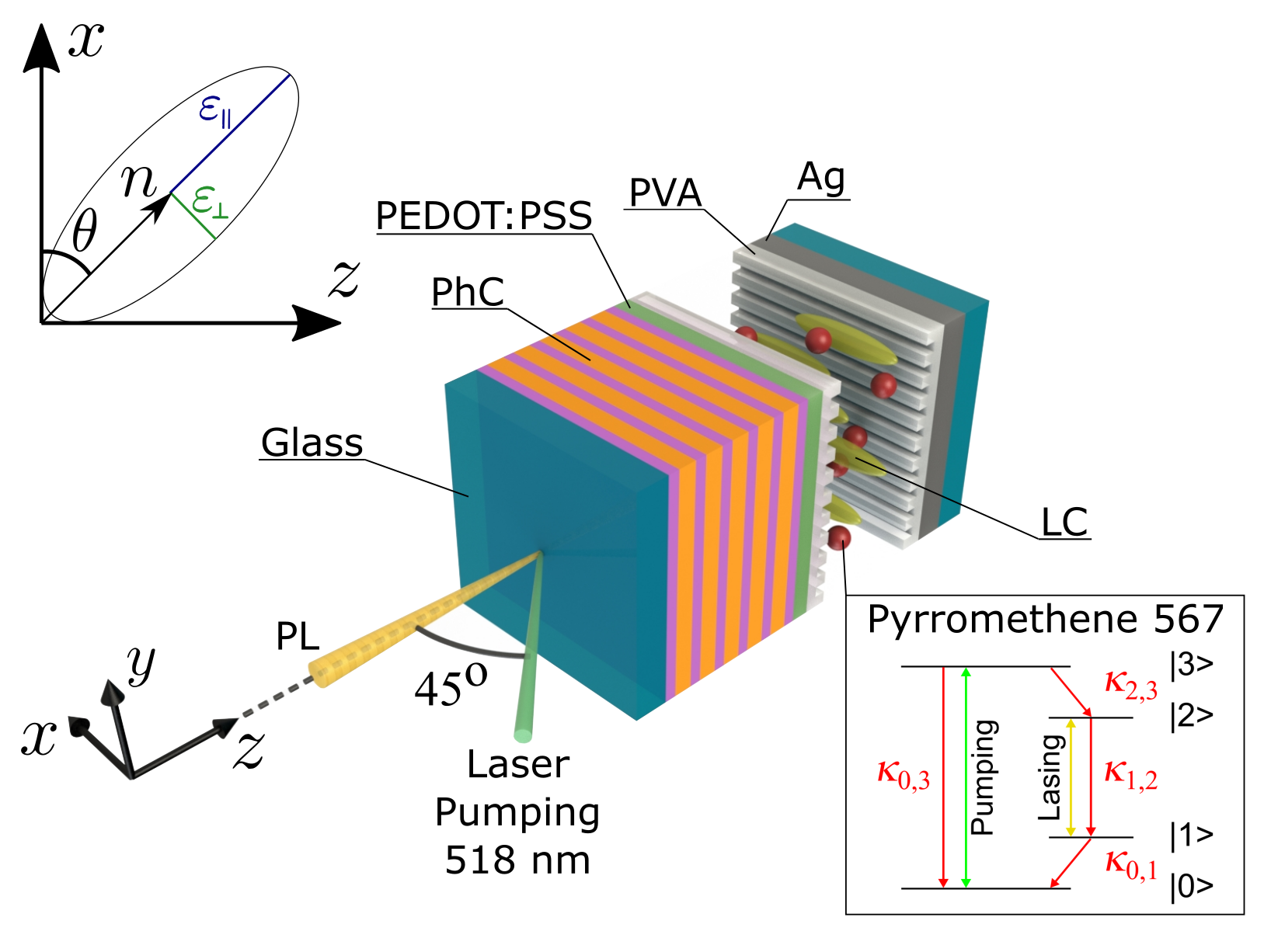}
    \caption{Microcavity model.
    The upper-left inset shows the LC director orientation and permittivity ellipse.
    The lower-right inset shows the four-level system for dye molecules.}
    \label{Fig1:Model}
\end{figure}

Our basic model is described by a three-term Hamiltonian
\begin{equation}
\widehat{\cal{H}}=\widehat{\cal{H}}_{ml}+
\widehat{\cal{H}}_f+
\widehat{\cal{H}}_c,
\end{equation}
where subscript $ml$ corresponds to molecular (emitter) subsystem, $f$ -- to the EM field, and $c$ -- to the coupling between the emitter and the EM field. The density matrix $\widehat{R}$ satisfies the Markovian master equation in the Lindblad form \cite{scully1997quantum}
\begin{equation}\label{CT_master1}
\frac{\partial \widehat{\cal{R}}}{\partial t}=\frac{1}{i\hbar}[\widehat{\cal{H}},\widehat{\cal{R}}]+\widehat{\cal{L}}_{\mathrm{nr}}(\widehat{\cal{R}})+\widehat{\cal{L}}_{\mathrm{r}}(\widehat{\cal{R}}),
\end{equation}
where the Liouvillian superoperator $\widehat{\cal{L}}_{\mathrm{nr}}(\widehat{\cal{R}})$ accounts for the non-radiative
intra-molecular transitions and acts only on the
emitter Hilbert space while $\widehat{\cal{L}}_{\mathrm{r}}(\widehat{\cal{R}})$ 
describes radiation of EM waves to the outer space and
acts on the EM field Hilbert space. It is shown in Appendix~\ref{LASER EQUATIONS} that Eq.~\ref{CT_master1} results in a set of coupled non-linear differential equations for dye polarizations, population densities and the electromagnetic fields
\begin{widetext}
\begin{align}\label{Full_Set1}
& \frac{\partial {{\bf J}}_{m}({\bf r})}{\partial t}+\kappa_{m}{\bf J}_{m}({\bf r}) = \varepsilon_0 \omega_{m}^2{\bf E}({\bf r}), \nonumber \\
& \frac{\partial{{\bf J}}_{p}({\bf r})}{\partial t}+\kappa_{p}{\bf J}_{p}({\bf r}) +
\bar{\omega}_{p}^2{\bf P}_{p}({\bf r})=-\frac{2\omega_{p}}{3\hbar}|{\bf d}_{p}|^2[\rho_3({\bf r})-\rho_0({\bf r})]{\bf E}({\bf r}), \nonumber \\
& \frac{\partial{{\bf J}}_{\ell}({\bf r})}{\partial t}+\kappa_{\ell}{\bf J}_{\ell}({\bf r})+
\bar{\omega}_{\ell}^2{\bf P}_{\ell}({\bf r})=-\frac{2\omega_{\ell}}{3\hbar}|{\bf d}_{l}|^2[\rho_2({\bf r})-\rho_1({\bf r})]{\bf E}({\bf r}), \nonumber \\
& \frac{\partial{\rho}_{3}({\bf r})}{\partial t}-\frac{1}{\hbar\omega_{p}}{\bf E}({\bf r})\cdot\left({\bf J}_{p}({\bf r}) + \frac{\kappa_{p}}{2}{\bf P}_{p}({\bf r})\right)
+
(\kappa_{0,3}+\kappa_{2,3}) \rho_3({\bf r})=
0, \nonumber \\
& \frac{\partial{\rho}_{2}({\bf r})}{\partial t}-\frac{1}{\hbar\omega_{\ell}}{\bf E}({\bf r})\cdot\left({\bf J}_{\ell}({\bf r})+\frac{\kappa_{\ell}}{2}{\bf P}_{\ell}({\bf r})\right)
+
\kappa_{1,2}\rho_2({\bf r})-\kappa_{2,3}\rho_3({\bf r})=0, \nonumber \\
& \frac{\partial{\rho}_{1}({\bf r})}{\partial t}+\frac{1}{\hbar\omega_{\ell}}{\bf E}({\bf r})\cdot\left({\bf J}_{\ell}({\bf r})+\frac{\kappa_{\ell}}{2}{\bf P}_{\ell}({\bf r})\right)
+
\kappa_{0,1} \rho_1({\bf r})-\kappa_{1,2}\rho_2({\bf r})=0, \nonumber \\
& \frac{\partial{\rho}_{0}({\bf r})}{\partial t}+\frac{1}{\hbar\omega_{p}}{\bf E}({\bf r})\cdot\left({\bf J}_{p}({\bf r})+\frac{\kappa_{p}}{2}{\bf P}_{p}({\bf r})\right)
-\kappa_{0,1}\rho_1({\bf r})-\kappa_{0,3}\rho_3({\bf r})=0, \nonumber \\
& \nabla\times{\bf E({\bf r})}=-\mu_0\frac{\partial {\bf H({\bf r})}}{\partial t}, \nonumber \\
& \nabla\times{\bf H}({\bf r})= \frac{\partial {\bf D}({\bf r})}{\partial t}, \nonumber \\
& {\bf E}({\bf r}) = \frac{ \widehat{\varepsilon}_h^{-1}({\bf r})}{\varepsilon_0}({\bf D}({\bf r}) - {\bf P}_{p}({\bf r}) - {\bf P}_{\ell}({\bf r}) - {\bf P}_{m}({\bf r})), \nonumber \\
& {\bf J}_n ({\bf r}) = \frac{\partial{\bf P}_n({\bf r})}{\partial t}, \; n = m, p , \ell,
\end{align}
\end{widetext}
where
\begin{align}
& \omega_p=\omega_3-\omega_0, \nonumber \\
& \omega_{\ell}=\omega_2-\omega_1, \nonumber \\
& \kappa_p=\kappa_{0,3}+\kappa_{2,3}, \nonumber \\
& \kappa_{\ell}=\kappa_{0,1}+\kappa_{1,2},\nonumber \\
& \bar{\omega}_{p, \ell}^2=\omega_{p, \ell}^2+\kappa_{p, \ell}^2/4.
\end{align}
Here $\bf E$ is the electric field strength, $\bf H$ is the magnetic field strength, $\bf D$ is the electric displacement field, $\bf P$ is the polarization,
$\bf J$ is the polarization current, $\bf d$ is the transition dipole moment.
The $\varepsilon_0$ is the vacuum permittivity, $\mu_0$ is the vacuum permeability, the $\widehat{\varepsilon}_h$ is the background permittivity tensor for host medium and the other materials of microcavity.
The subscript $m$ stands for metal free electrons in the Drude model, i.e. $\kappa_m$ is the damping rate and $\omega_{m}$ is the plasma frequency.
The subscript $p$ stands for transition between 0-th and 3-rd levels, the subscript $\ell$ stands for transition between 1-st and 2-nd levels. %
The $\kappa_{i,j}$ stands for the non-radiative decay rate between the $j$-th and $i$-th levels, $\omega_j$ stands for the frequency of the $j$-th  level, $\rho_j$ is the $j$-th' energy level population density ($j = 0,1,2,3$), see inset in Fig.~\ref{Fig1:Model}.

\section{LC MICROCAVITY SET-UP}

The microcavity is composed of PhC and silver (Ag) mirrors deposited on glass substrates, see Fig.~\ref{Fig1:Model}.
The PhC mirror comprises five $Si_{3}N_{4}$ and $SiO_{2}$ bilayers with an additional $Si_{3}N_{4}$ layer placed on top of the structure.
The thickness of the $Si_{3}N_{4}$ and $SiO_{2}$ layers is 59 and 87~nm, respectively.
The PhC mirror is covered by a conducting layer of poly(3,4-ethylenedioxythiophene)-poly(styrene-
sulfonate) (PEDOT:PSS) (\textit{Sigma Aldrich}).
The microcavity is filled by nematic LC 4-pentyl-4’-cyanobiphenyl (5CB) (\textit{Belarusian State Technological University}) doped with dye Pyrromethene 567 (\textit{Exciton}) at a weight concentration of 0.43$\%$.
In order to induce the planar alignment of the liquid crystal, both mirrors are coated with poly(vinyl alcohol) (PVA) (\textit{Sigma Aldrich}). The polymer films are unidirectionally rubbed.
The thickness of the liquid crystal layer $L$ is equal to 3.5~$\mu$m.

\begin{figure}
    \centering    \includegraphics[width=1\linewidth]{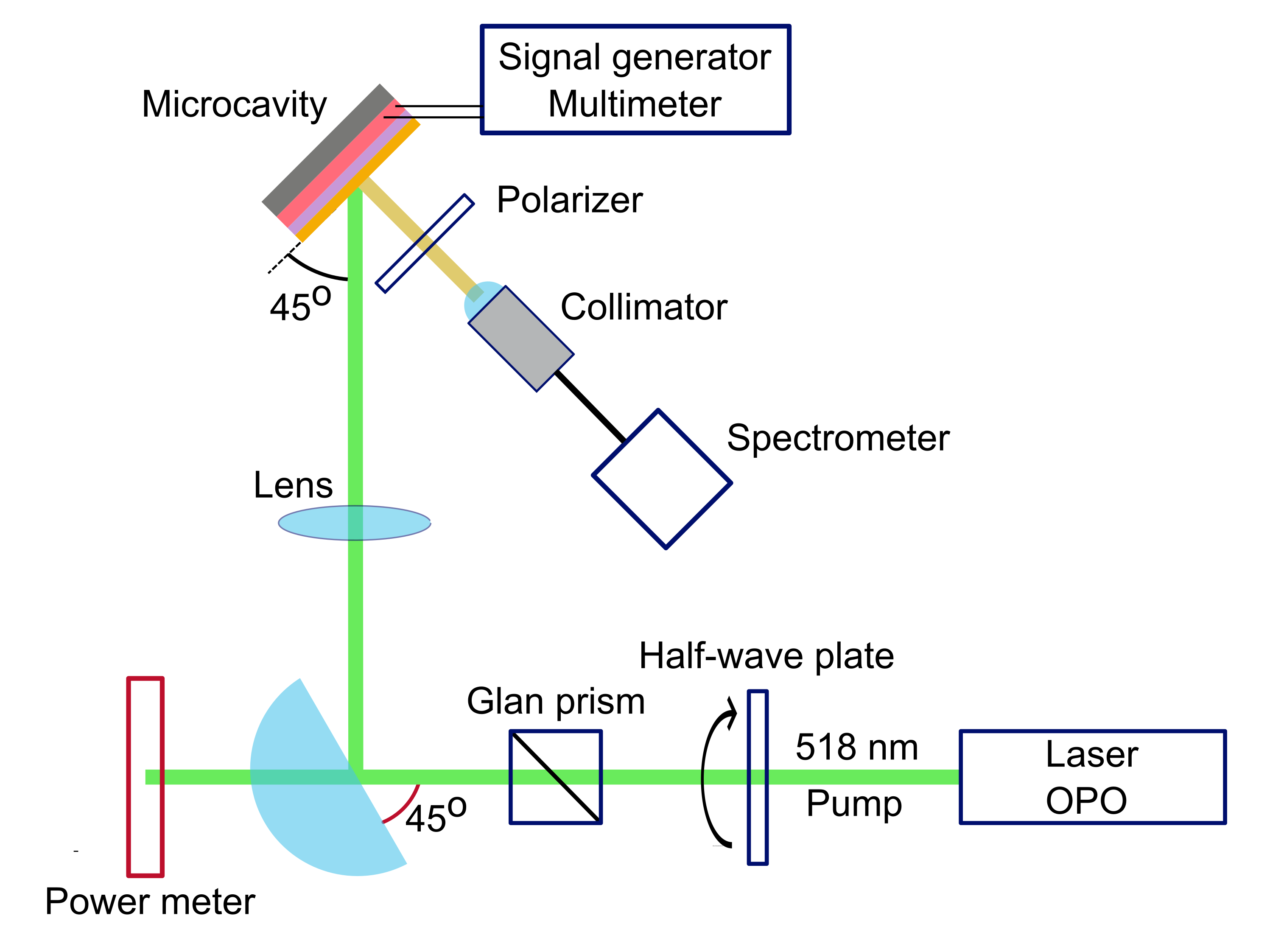}    \caption{Experimental set-up}    \label{Fig2:Setup}
\end{figure}

The transmittance spectra of the PhC at normal incidence and reflectance spectra of the sample at $8^{\circ}$ were measured with a \textit{Thorlabs OSL2} incoherent source and a fiber-optic spectrometer \textit{Ocean FX-UV-VIS}. The opaque \textit{Thorlabs} Ag mirror was used as the reference for reflectance spectra measurements. 
The experimental setup for photoluminescence (PL) measurements is schematically shown in Fig.~\ref{Fig2:Setup}.
The radiation from \textit{LOTIS-TII} optical 
parametric oscillator (OPO) was used for pumping of the microcavity.
The OPO itself was pumped by a third-harmonic from a Q-switched Nd:YAG pulse laser at wavelength 355~nm with a duration of 13~ns and a repetition rate of 10~Hz. 
The pulse energy was controlled by a half-wave plate mounted on a \textit{Thorlabs KPRM1E} rotated platform and Glan prism.
The transmitted through a beam splitter light was incident on a \textit{Ophir Nova II} power meter, while the reflected part was focused with a 10~cm-long focal length lens onto the microcavity at a $45^{\circ}$ incident angle in order to discern the pumping beam and the PL signal during the collection of data.
The polarization of a collected PL signal was controlled by a polarizer.
The PL spectra were obtained with a collimator integrated with a fiber-optic spectrometer \textit{Ocean FX-UV-VIS}.
A signal generator \textit{Aktakom AWG-4150} provided 1 kHz square-wave ac
voltage $U$ applied to PEDOT:PSS and Ag layers. 

\section{RESULTS AND DISCUSSION}

\begin{figure}
    \centering    \includegraphics[width=1\linewidth]{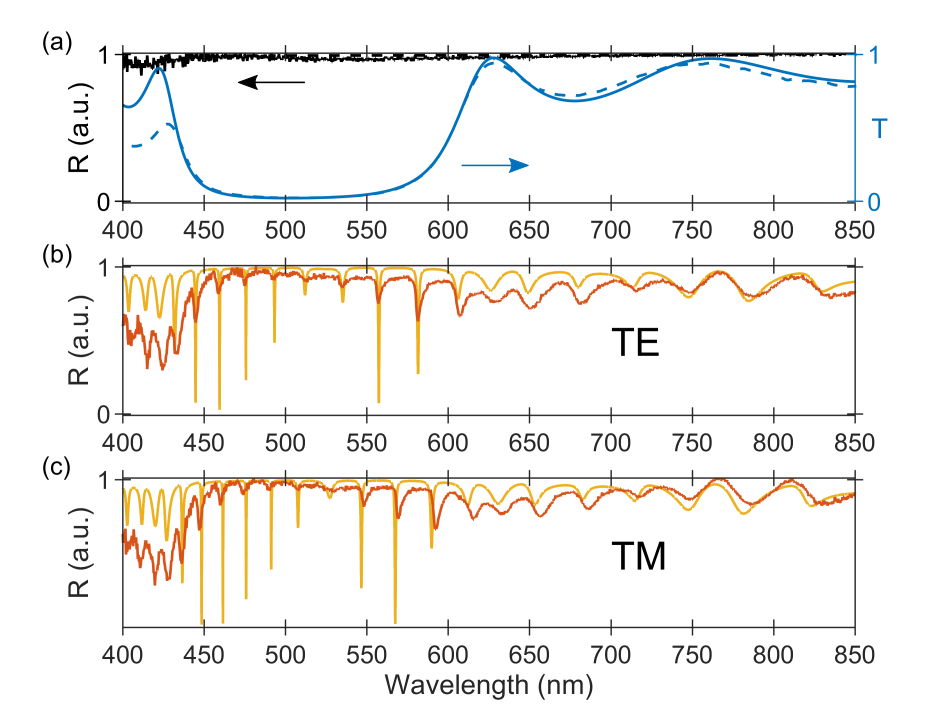}
    \caption{(a) The measured (dashed blue line) and calculated (solid blue line)  transmittance spectra of the PhC at normal incidence.
    The measured (dashed black line) and calculated (solid black line)  reflectance spectra of the Ag mirror for TE-polarized light incident at $8^{\circ}$. (b,c) The measured (red line) and calculated (orange line) reflectance spectra of the microcavity for TE-(b) and TM-polarized (c) light incident at $8^{\circ}$. The LC director polar angle $\theta = 0.9^\circ$ = 
 const(z).} \label{Fig3:Spectra}
\end{figure}

The transmittance spectrum of the PhC at normal incidence is shown in Fig.~\ref{Fig3:Spectra}(a).
The spectral range from 450~nm to 600~nm with a low transmittance corresponds to a photonic band gap.
The transmittance of the Ag mirror in the above-mentioned spectral range is zero, while the reflectance is close to a reflectance of the 
{\it Thorlabs} opaque mirror, see Fig.~\ref{Fig3:Spectra}(a).

The reflectance spectra of the microcavity demonstrate numerous dips corresponding to excitation of microcavity modes, see Fig.~\ref{Fig3:Spectra}~(b,c).
Since the LC layer is anisotropic, the number and the positions of the resonant lines are different for TE and TM-polarized light.
It can be seen, that amplitude of resonant lines is lower near the wavelength 524~nm, due to the light absorption by dye molecules.

\begin{figure}[t]
    \center{   \includegraphics[width=1\linewidth]{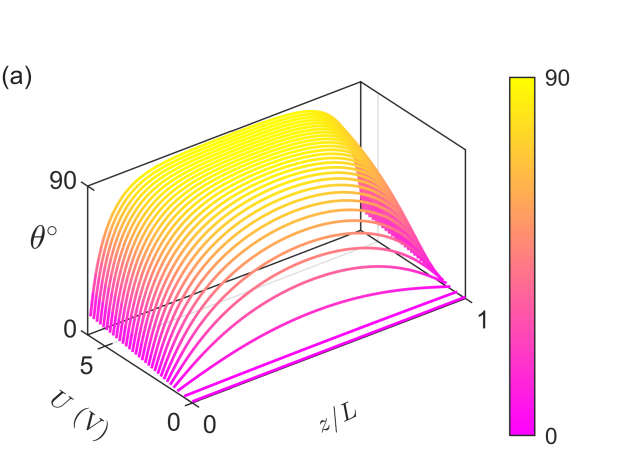}\\
    \includegraphics[width=1\linewidth]{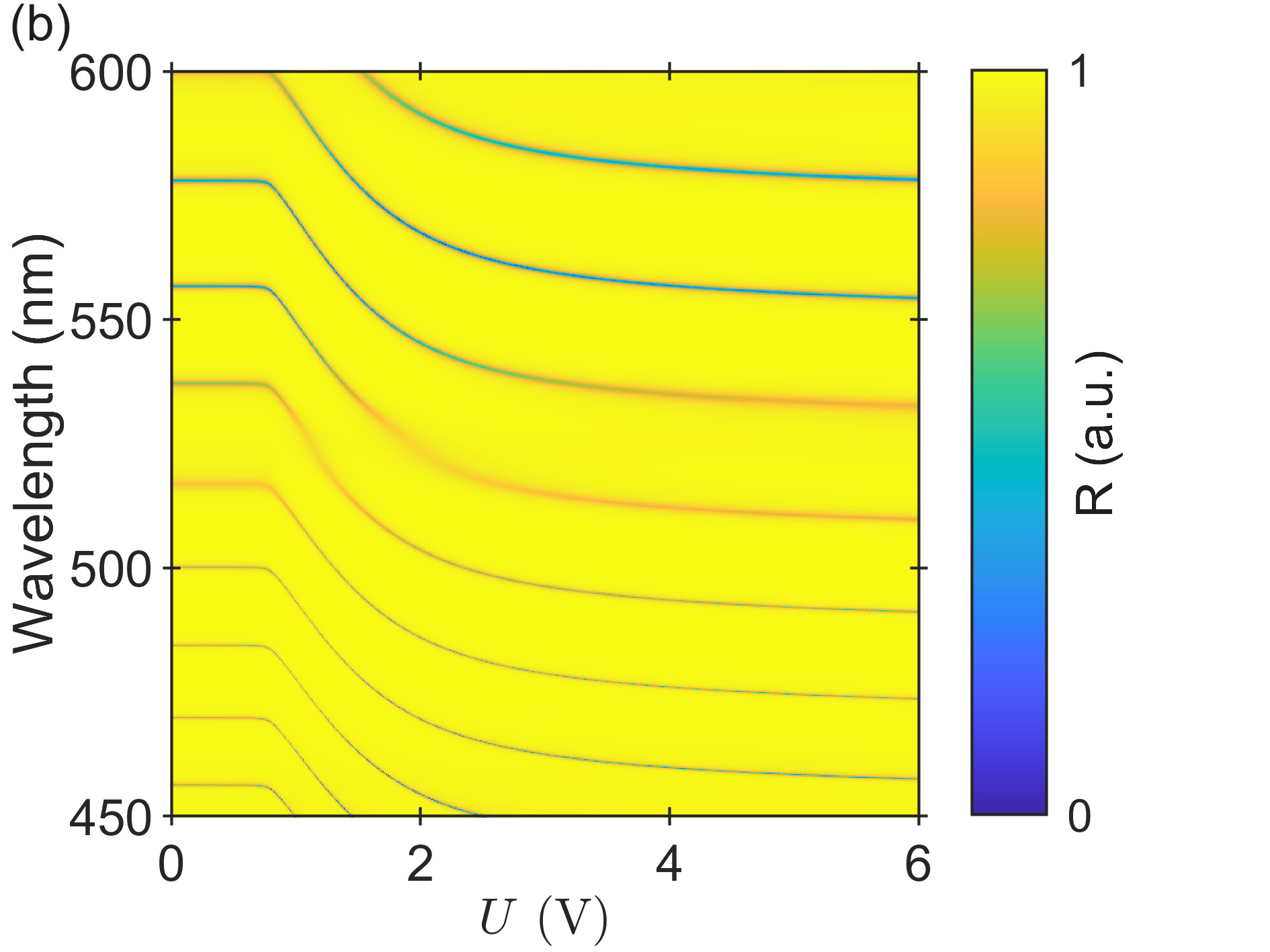}
    \caption{(a) The calculated distribution $\theta(z,U)$.
    (b) The calculated reflectance spectrum for $x$-polarized light at normal incidence.}
    }
    \label{Fig4:LC}
\end{figure}

The measured spectra are superimposed with the calculated ones, see Fig.~\ref{Fig3:Spectra}.
The spectra were calculated using the scattering matrix method~\cite{rumpf2011improved}, taking into account the frequency dispersion of permittivities~\cite{rodriguez2016self, luke2015broadband,chen2015optical,bodurov2016modified,french2009optical}.
The Ag layer was described by the Drude model Eq.~\eqref{Drude_Response}, the LC layer doped with dye molecules was described by the Lorentz model Eq.~\eqref{Lorentz_Response}, see Appendix~\ref{LINEAR RESPONSE}.
The agreement in spectral positions of measured and calculated resonant lines can be seen.
The measured resonant lines are broadened due to a nonuniform LC layer thickness over a light spot of 5 mm diameter.

The applied external voltage $U$ induces the Frederiks effect in the LC layer, i.e. the  director $\boldsymbol{n}$ of the investigated LC tends to align with the external field direction~\cite{blinov2010structure}.
Figure~\ref{Fig4:LC}(a) shows the calculated distribution $\theta(z, U)$ for polar angle of LC director orientation, see inset in Fig.~\ref{Fig1:Model}.
The distribution $\theta(z,U)$ was calculated using the Frank-Oseen model, taking into account the finite value of a surface anchoring energy given by Rapini potential, see Appendix~\ref{FRANK-OSEEN MODEL}.
The initial homogeneous distribution $\theta(z) = 0.9^{\circ}$ at $U = 0$~V is enforced by pretilt due to the substrates rubbing. 
It can be seen, that after the Frederiks threshold voltage $\sim 0.75$~V angle $\theta$ tends to 90$^{\circ}$ in the whole LC volume.
Moreover, due to the finite surface anchoring energy the value of $\theta$ on the substrates  increases up to $12.7^{\circ}$ at $U = 6$~V~\cite{jiao2008alignment}.

\begin{figure}
    \centering    \includegraphics[width=1\linewidth]{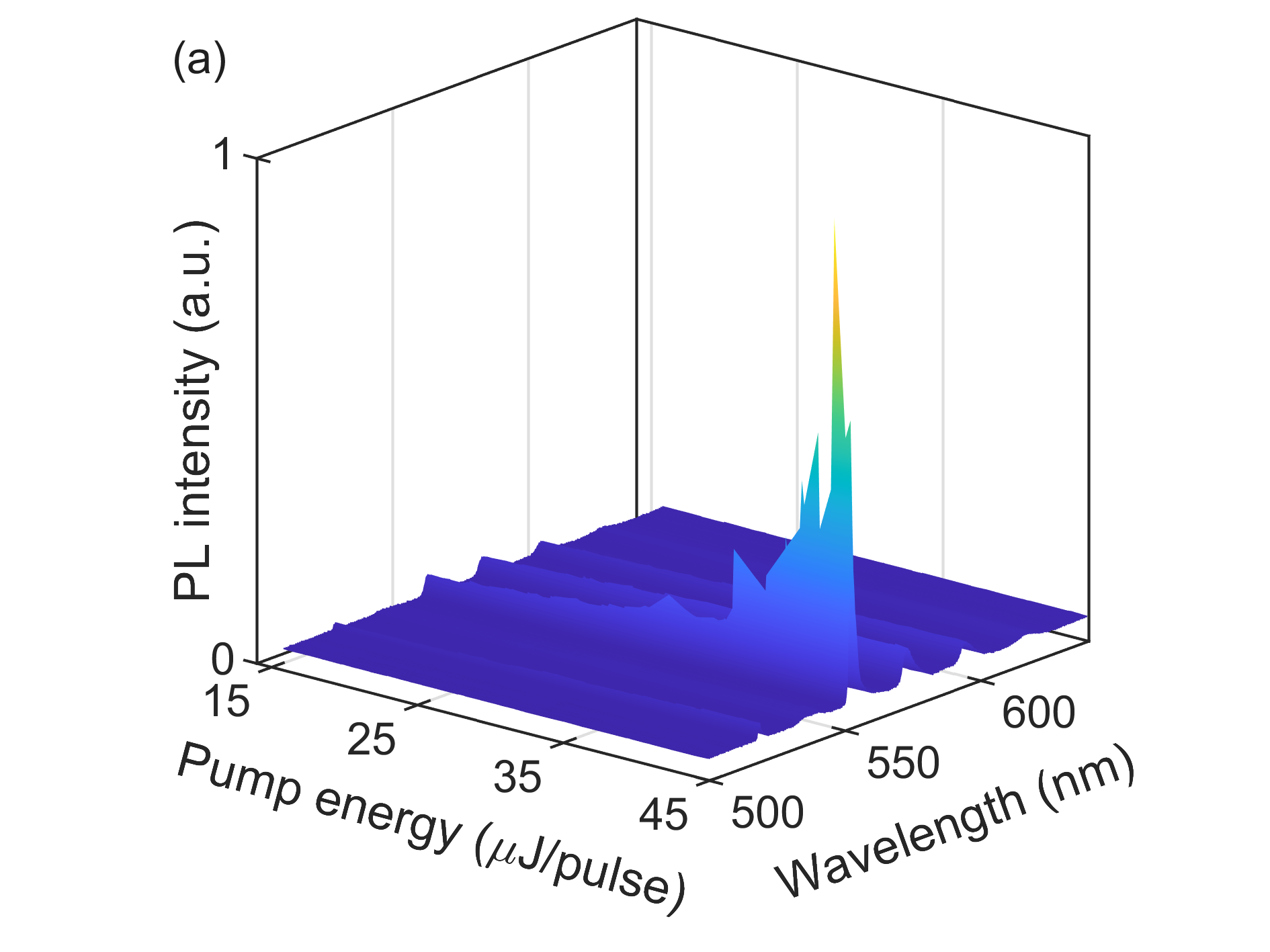}  \includegraphics[width=1\linewidth]{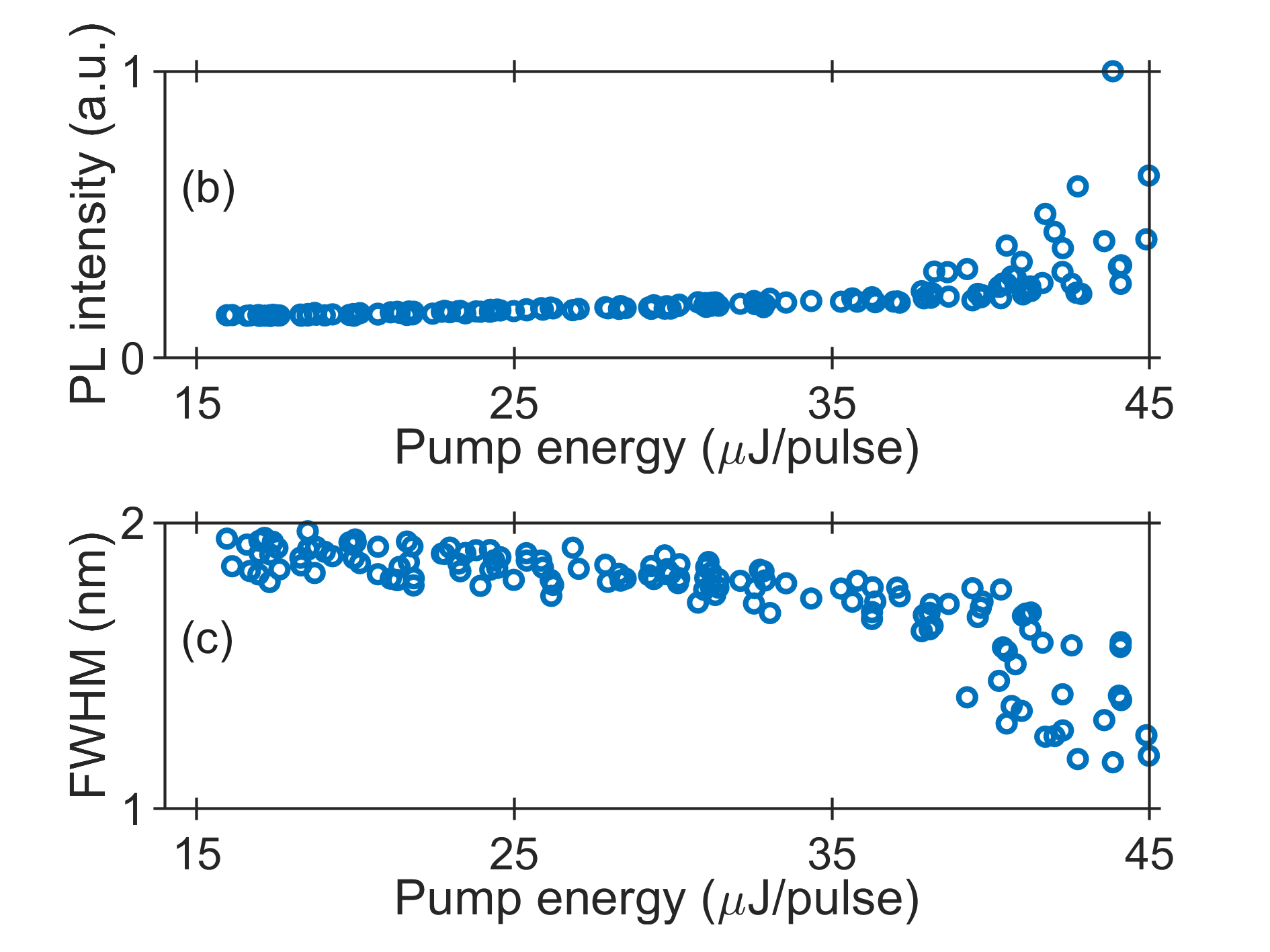}
    \caption{(a) The measured dependence of PL spectra on pumping energy.
    (b) PL peak intensity and peak linewidth defined as full width at half maximum (FWHM) at wavelength 552~nm. Applied voltage $U = 0$~V.}
    \label{Fig5:Pump}
\end{figure}

Figure~\ref{Fig4:LC}(b) shows the calculated reflectance spectrum for $x$-polarized light at normal incidence. 
The change in $\theta(z,U)$ leads to a variation in the LC permittivity tensor Eq.~\eqref{LC_epsilon}.
In more detail, $\theta$ change from $0.9^{\circ}$ to $90^{\circ}$ leads to the  $\varepsilon_{xx}$ change from $\varepsilon_{\parallel}$ to $\varepsilon_{\perp}$.
The variation of the permittivity tensor reveals in the blue shift of resonant lines after the Frederiks threshold voltage, as seen in Fig.~\ref{Fig4:LC}(b).
The amplitude of resonant lines drops down close to wavelength of 524~nm corresponding to the absorption maxima for the dye molecules.   

Figure~\ref{Fig5:Pump}(a) shows the measured PL spectrum under pumping at wavelength 518~nm and applied voltage $U = 0$~V.
The PL spectrum demonstrates the resonant peaks corresponding to excitation of the microcavity modes.
The increase in pumping energy leads to an increase in the PL signal intensity.
It can be seen, that one of the microcavity modes at wavelength 552~nm survives in mode competition above the laser threshold pumping energy $\sim 35\mu$J/pulse.
Above the laser threshold the amplitude of the PL intensity signal drastically increases, while the linewidth decreases, see Fig.~\ref{Fig5:Pump}(b)~\cite{ozaki2003electrically}.

\begin{figure}
    \centering    \includegraphics[width=1\linewidth]{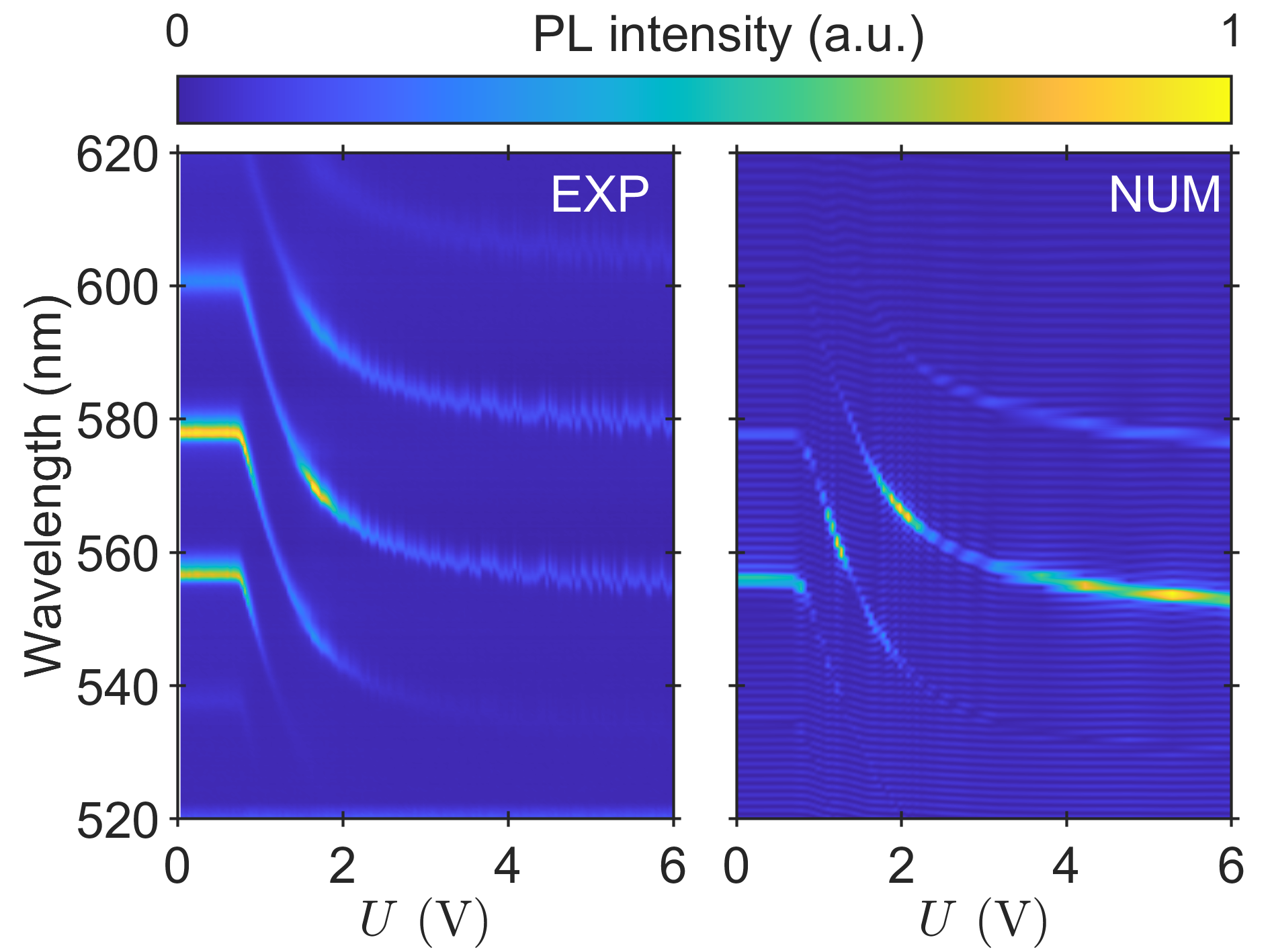}
    \caption{The measured (a) and calculated (b) PL spectra under applied voltage $U$.}
    \label{Fig6:PL}
\end{figure}

The measured and calculated PL spectra above the laser threshold are shown in Fig.~\ref{Fig6:PL}.
The change in position and amplitude of the lasing peak under an applied voltage can be seen.
The shift in position of lasing peak is consistent 
with the shift in position of microcavity mode resonant line in the reflectance spectrum, see Fig.~\ref{Fig4:LC}(b).
The microcavity was being pumped for all time voltage was applied, which led to the dye bleaching~\cite{morris2005effects, huang2006lasing}.
As the result, the PL amplitude intensity decrease above the voltage of $\sim 2.5$~V, see Fig.~\ref{Fig6:PL}(a).
The discontinuous change of the lasing wavelength is observed in both, measured and calculated spectra.
This phenomenon was observed in the first realization of the PhC/LC/PhC microcavity laser~\cite{ozaki2003electrically} and  can be explained by the mode competition, beacause the mode which is closer to the dye emission band maxima has the maximum gain.

\section{CONCLUSION}

The comprehensive approach for simulating lasing dynamics in liquid crystal based systems was presented.
The transformation of the liquid crystal host structure under an applied voltage was calculated in the framework of the Frank-Oseen model with the finite liquid crystal anchoring energy described by the Rapini potential.
The corresponding change in the liquid crystal permittivity was taken into account in Maxwell's equations, which were solved  consistently with the equations for the emitters' level population and polarizations.
The emitters were described as four-level systems with two radiative transitions.
The equations for emitters' energy level population dynamics were derived from the Markovian master equation in the Lindblad form, which took into account the non-radiative transitions.
The final set of 
equations contained the non-radiative decay rates only, while dynamics of radiative transitions was governed by the solution of equations.
The system of differential equations was solved numerically by using the finite difference time domain method on the staggered Yee grid.
The theoretical results agree with the experimentally measured electrically controlled lasing spectra
of the microlaser.
The demonstrated microlaser consists of photonic crystal and silver mirrors with the resonant layer made of
liquid crystal doped with dye molecules.
Our approach paves the way for novel cholesteric liquid crystal lasers
and liquid crystal driven vertical cavity surface emitting laser arrays with active electrically controlled beam steering.

\section*{Author Contributions}
D.S.B. -- Investigation, Visualization, Writing – review and editing.
P.S.P. -- Investigation,
Methodology, Software, Formal analysis, Writing – original draft.
D.N.M. --Investigation,
Methodology, Software, Writing – review and editing.
V.S.S. --Investigation,
Methodology, Writing – review and editing.
G.A.R., R.G.B., S.V.N., N.A.Z., I.A.T.   -- Investigation, Writing – review and editing.
S.Ya.V. -- Conceptualization, Writing – review and editing.
K-P.C. -- Funding acquisition, Supervision, Writing – review and editing.
I.V.T. -- Funding acquisition, Supervision, Writing – review and editing.

\section*{Conflicts of interest}
The authors declare no conflicts of interest.

\section*{Data Availability Statement}
The data that support the findings of this study are presented in the Appendix section.

\section*{ACKNOWLEDGMENTS}

This research was funded by the Russian Science Foundation (project no. 22-42-08003). This work was supported by the National Science and Technology Council (NSTC 112-2223-E-007-007-MY3; 111-2923-E-007 -008 -MY3; 111-2628-E-007-021.)
The authors would like to express their special thanks to Krasnoyarsk Regional Center of Research Equipment of Federal Research Center "Krasnoyarsk Science Center SB RAS" for providing equipment to ensure the accomplishment of this project.

\newpage

\appendix

\begin{widetext}

\newpage

\section{LASER EQUATIONS}
\label{LASER EQUATIONS}

We start with a three-term Hamiltonian
\begin{equation}
\widehat{\cal{H}}=\widehat{\cal{H}}_{ml}+
\widehat{\cal{H}}_f+
\widehat{\cal{H}}_c,
\end{equation}
where subscript $ml$ stands for molecular (emitter) subsystem, $f$ -- for the EM field, and $c$ -- for the coupling between the emitter and the EM field. We expect the density matrix
to satisfy the Lindblad master equation in the following form
\begin{equation}\label{CT_master}
\frac{\partial \widehat{\cal{R}}}{\partial t}=\frac{1}{i\hbar}[\widehat{\cal{H}},\widehat{\cal{R}}]+\widehat{\cal{L}}_{\mathrm{nr}}(\widehat{\cal{R}})+\widehat{\cal{L}}_{\mathrm{r}}(\widehat{\cal{R}}),
\end{equation}
where the Liouvillian superoperator $\widehat{\cal{L}}_{\mathrm{nr}}(\widehat{\cal{R}})$ accounts for the non-radiative
intra-molecular transitions and acts only on the
emitter Hilbert space while $\widehat{\cal{L}}_{\mathrm{r}}(\widehat{\cal{R}})$ 
describes radiation of EM waves to the outer space and
acts on the EM field Hilbert space.
In a four-level laser the molecular subsystem can
be found in four states such that $\hbar\omega_0<\hbar\omega_1<\hbar\omega_2<\hbar\omega_3$.
It is assumed that there are non-zero dipole momenta of intramolecular transitions $|0\rangle\rightarrow|3\rangle$ and $|1\rangle\rightarrow |2\rangle$ which are termed ${\bf d}_{3,0}$ and ${\bf d}_{2,1}$, correspondingly.
The other radiative transitions are not allowed in the system.
The polarizations associated with the allowed radiation transitions will be denoted by the same subscripts.
The populations of the levels is described by the population densities $\rho_n({\bf r}), \ n=0,1,2,3$ such as
\begin{equation}
\rho_0({\bf r})+\rho_1({\bf r})+\rho_2({\bf r})+\rho_3({\bf r})=\rho({\bf r})
\end{equation}
$\rho({\bf r})$ being the density of emitters.
The molecular part of the Hamiltonian reads
\begin{equation}
\widehat{\cal H}_{\rm ml}=\sum_{m=0}^3\hbar\omega_{m}\widehat{X}_{m,m},
\end{equation}
where
\begin{equation}
\widehat{X}_{m,m'}=|m\rangle\langle m'|,
\end{equation}
is the Hubbard operator.
For a single emitter 
the light-matter interaction is described 
by the dipole Hamiltonian
\begin{equation}
\widehat{\cal H}_{\rm c}=-({\bf d}_{3,0}\widehat{X}_{3,0}+
{\bf d}_{3,0}^*\widehat{X}_{0,3})\cdot
\widehat{ {\bf E}}({\bf r}_0)
-({\bf d}_{2,1}\widehat{X}_{2,1}+
{\bf d}_{2,1}^*\widehat{X}_{1,2})\cdot
\widehat{ {\bf E}}({\bf r}_0),
\end{equation}
where $\widehat{ {\bf E}}({\bf r}_0)$ is the field operator in the position of the emitter ${\bf r}_0$.
The relaxation operator for the non-radiative transitions reads
\begin{equation}
\widehat{\cal L}_{nr}({\widehat{\cal R}})=-\frac{1}{2}
\sum_{m=0}^4\sum_{m'=0}^{m-1}\kappa_{m,m'}(\widehat{X}_{m'm}\widehat{X}_{m'm}{\widehat{\cal R}}
-2\widehat{X}_{m'm}{\widehat{\cal R}}\widehat{X}_{m'm}
+{\widehat{\cal R}}\widehat{X}_{m'm}\widehat{X}_{m'm}).
\end{equation}
In the above formula the summation over $m'$ is run up to $m-1$ which means that the possibilities of transverse relaxation and incoherent pumping are not taken into account.
Finally, as it will be clear later on we do not need the mathematical expressions for $\widehat{\cal{H}}_f$ and $\widehat{\cal{L}}_{\mathrm{r}}(\widehat{\cal{R}})$.

To derive the dynamic equations in the coherent approximation we define the following quantities
\begin{align}\label{app00}
&\rho_n=\sigma_{n,n}=\langle \widehat{X}_{n,n}\rangle ,\nonumber \\
&\sigma_{n,n'}=\langle \widehat{X}_{n,n'}\rangle, \ \ n\neq n', \nonumber \\
&\sigma_{n',n}=\sigma_{n,n'}^*.
\end{align}
The dynamic equations for the above quantities can be obtained by applying their definitions to the master equation Eq.~\eqref{CT_master}.
The following formulas are used to facilitate the calculations
\begin{align}
&\frac{1}{i\hbar}{\mathrm tr}(\widehat{X}_{n,n'}[\widehat{\cal H}_{\rm ml},
\widehat{\cal R}])=i(\omega_n-\omega_{n'})\sigma_{n,n'} ,\nonumber \\
&\frac{1}{i\hbar}{\mathrm tr}(\widehat{X}_{n,n'}[\widehat{\cal H}_{\rm c},
\widehat{\cal R}])=-\frac{1}{i\hbar}
\sum_m {\bf E}({\bf r})\cdot\left[ ({\bf d}_{n',m}\sigma_{n,m} -{\bf d}_{m,n}\sigma_{m,n'})+
({\bf d}^*_{m,n'}\sigma_{n,m} -{\bf d}^*_{n,m}\sigma_{m,n'})
\right], \nonumber \\
&{\mathrm tr}(\widehat{X}_{n,n'}\widehat{\cal L}(\widehat{\cal R}))=\sum_m\kappa_{n,m}\delta_{n,n'}\sigma_{n,n}-\frac{1}{2}\sum_m(\kappa_{m,n'}+\kappa_{m,n})\sigma_{n,n'},
\end{align}
where
$ { {\bf E}}({\bf r})=\langle\widehat{ {\bf E}}({\bf r})\rangle $.
The above leads to the following dynamic equations 
\begin{align}\label{app0}
&\dot{\rho_0}=\frac{1}{i\hbar}({\bf E}({\bf r})\cdot{\bf d}_{3,0}\sigma_{3,0}-{\bf E}({\bf r})\cdot{\bf d}_{3,0}^*\sigma_{0,3})+\kappa_{0,1}\rho_1+\kappa_{0,3}\rho_3, \nonumber \\
&\dot{\rho_1}=\frac{1}{i\hbar}({\bf E}({\bf r})\cdot{\bf d}_{2,1}\sigma_{2,1}-{\bf E}({\bf r})\cdot{\bf d}_{2,1}^*\sigma_{1,2})
+\kappa_{1,2}\rho_2-\kappa_{0,1}\rho_1,
\nonumber \\
&\dot{\rho_2}=-\frac{1}{i\hbar}({\bf E}({\bf r})\cdot{\bf d}_{2,1}\sigma_{2,1}-{\bf E}({\bf r})\cdot{\bf d}_{2,1}^*\sigma_{1,2})
+\kappa_{2,3}\rho_3-\kappa_{1,2}\rho_2,
\nonumber \\
&\dot{\rho_3}=-\frac{1}{i\hbar}({\bf E}({\bf r})\cdot{\bf d}_{3,0}\sigma_{3,0}-{\bf E}({\bf r})\cdot{\bf d}_{3,0}^*\sigma_{0,3})
-\kappa_{0,3}\rho_3-\kappa_{2,3}\rho_3,
\nonumber \\
&\dot{\sigma}_{3,0}=i(\omega_3-\omega_0)\sigma_{3,0}-\frac{\kappa_{0,3}+\kappa_{2,3}}{2}\sigma_{3,0}
-\frac{1}{i\hbar}({\bf E}({\bf r})\cdot{\bf d}^*_{3,0}\sigma_{3,3}-{\bf E}({\bf r})\cdot{\bf d}_{3,0}^*\sigma_{0,0}), \nonumber \\
&\dot{\sigma}_{2,1}=i(\omega_2-\omega_1)\sigma_{2,1}-\frac{\kappa_{0,1}+\kappa_{1,2}}{2}\sigma_{2,1}-\frac{1}{i\hbar}({\bf E}({\bf r})\cdot{\bf d}^*_{2,1}\sigma_{2,2}-{\bf E}({\bf r})\cdot{\bf d}_{2,1}^*\sigma_{1,1}).
\end{align}
These equations do not constitute a full set since equations for the coherent electric field strength ${\bf E}({\bf r})$ are missing.
To obtain a closed set of equations, the system  Eq.~\eqref{app0} has to be complemented with two of Maxwell's equations, namely Ampere-Maxwell and Faraday's laws. This will be done in the end of the section.

To proceed we introduce some new definitions. First, we denote the frequency differences by
\begin{align}\label{omega}
& \omega_p=\omega_3-\omega_0, \nonumber \\
& \omega_{\ell}=\omega_2-\omega_1,
\end{align}
where $\omega_p$ is the frequency of the short-wavelength (pumping) transition while $\omega_{\ell}$ is the frequency of the long-wavelength (lasing) transition. In the same manner for the corresponding off-diagonal decay rates we write
\begin{align}\label{kappa}
& \kappa_p=\kappa_{0,3}+\kappa_{2,3}, \nonumber \\
&\kappa_{\ell}=\kappa_{0,1}+\kappa_{1,2}.
 \end{align}
The dipole moments of both transitions can be calculated as
\begin{align}\label{app1}
{\bf d}_{p}={\bf d}_{3,0}\sigma_{3,0}+{\bf d}^*_{3,0}\sigma_{0,3}, \nonumber \\
{\bf d}_{\ell}={\bf d}_{2,1}\sigma_{2,1}+{\bf d}^*_{2,1}\sigma_{1,2}.
\end{align}
In the case of $N$ distributed emitters the dipole moments are replaced by polarizations
\begin{equation}
{\bf P}_{p,\ell}({\bf r})=\sum_n^N {\bf d}_{p,\ell}\delta({\bf r}-{\bf r}_n)
\end{equation}
where ${\bf r}_n$ are the coordinates of emitters. By using Eq.~\eqref{app0} together with Eq.~\eqref{app1} and averaging over random orientations of emitters we find
\begin{align}\label{Four_level}
& \frac{\partial^2{{\bf P}}_{p}({\bf r})}{\partial t^2}+\kappa_{p}\frac{\partial{\bf P}_{p}({\bf r})}{\partial t}+
\bar{\omega}_{p}^2{\bf P}_{p}({\bf r})=-\frac{2\omega_{p}}{3\hbar}|{\bf d}_{p}|^2[\rho_3({\bf r})-\rho_0({\bf r})]{\bf E}({\bf r}), \nonumber \\
& \frac{\partial^2{{\bf P}}_{\ell}({\bf r})}{\partial t^2}+\kappa_{\ell}\frac{\partial{\bf P}_{\ell}({\bf r})}{\partial t}+
\bar{\omega}_{\ell}^2{\bf P}_{\ell}({\bf r})=-\frac{2\omega_{\ell}}{3\hbar}|{\bf d}_{l}|^2[\rho_2({\bf r})-\rho_1({\bf r})]{\bf E}({\bf r}), \nonumber \\
& \frac{\partial{\rho}_{3}({\bf r})}{\partial t}-\frac{1}{\hbar\omega_{p}}{\bf E}({\bf r})\cdot\left(\frac{\partial{{\bf P}}_{p}({\bf r})}{\partial t}+\frac{\kappa_{p}}{2}{\bf P}_{p}({\bf r})\right)
+
(\kappa_{0,3}+\kappa_{2,3}) \rho_3({\bf r})=
0, \nonumber \\
& \frac{\partial{\rho}_{2}({\bf r})}{\partial t}-\frac{1}{\hbar\omega_{\ell}}{\bf E}({\bf r})\cdot\left(\frac{\partial{{\bf P}}_{\ell}({\bf r})}{\partial t}+\frac{\kappa_{\ell}}{2}{\bf P}_{\ell}({\bf r})\right)
+
\kappa_{1,2}\rho_2({\bf r})-\kappa_{2,3}\rho_3({\bf r})=0, \nonumber \\
& \frac{\partial{\rho}_{1}({\bf r})}{\partial t}+\frac{1}{\hbar\omega_{\ell}}{\bf E}({\bf r})\cdot\left(\frac{\partial{{\bf P}}_{\ell}({\bf r})}{\partial t}+\frac{\kappa_{\ell}}{2}{\bf P}_{\ell}({\bf r})\right)
+
\kappa_{0,1} \rho_1({\bf r})-\kappa_{1,2}\rho_2({\bf r})=0, \nonumber \\
& \frac{\partial{\rho}_{0}({\bf r})}{\partial t}+\frac{1}{\hbar\omega_{p}}{\bf E}({\bf r})\cdot\left(\frac{\partial{{\bf P}}_{p}({\bf r})}{\partial t}+\frac{\kappa_{p}}{2}{\bf P}_{p}({\bf r})\right)
-\kappa_{0,1}\rho_1({\bf r})-\kappa_{0,3}\rho_3({\bf r})=0, \nonumber \\
& \nabla\times{\bf E({\bf r})}=-\mu_0\frac{\partial {\bf H({\bf r})}}{\partial t}, \nonumber \\
& \nabla\times{\bf H}({\bf r})=\varepsilon_0\widehat{\varepsilon}_h({\bf r})\frac{\partial {\bf E}({\bf r})}{\partial t}+\frac{\partial {\bf P}_{p}({\bf r})}{\partial t}
+\frac{\partial {\bf P}_{\ell}({\bf r})}{\partial t},
\end{align}
where 
\begin{align}
\bar{\omega}_{n}^2=\omega_{n}^2+\frac{\kappa_{n}^2}{4}, \ \ n = p, \ell.
\end{align}
Note that the last two lines in Eq.~\eqref{Four_level} are Maxwell's equation for the EM field, $\widehat{\varepsilon}_h ({\bf r})$ being the background permittivity tensor for host medium and the other materials the microcavity is made of.

Since the microcavity includes the metal mirror, the system of equations~Eq.\eqref{Four_level} has to be complemented with the equation for polarization  ${\bf P}_{m}$ of free electrons in the Drude model
\begin{equation}
\label{Drude}
\frac{\partial^2{{\bf P}}_{m}({\bf r})}{\partial t^2}+\kappa_{m}\frac{\partial{\bf P}_{m}({\bf r})}{\partial t} = \varepsilon_0 \omega_{m}^2{\bf E}({\bf r}),
\end{equation}
where $\kappa_{m}$ is the damping rate and $\omega_{m}$ is the plasma frequency.
The last line of Eq.~\eqref{Four_level},
then, reads as
\begin{equation}\label{H}
     \nabla\times{\bf H}({\bf r})=\varepsilon_0\widehat{\varepsilon}_h ({\bf r}) \frac{\partial {\bf E}({\bf r})}{\partial t}+\frac{\partial {\bf P}_{p}({\bf r})}{\partial t}
+\frac{\partial {\bf P}_{\ell}({\bf r})}{\partial t} + \frac{\partial {\bf P}_{m}({\bf r})}{\partial t}
\end{equation}
To simplify the above equation we use the electric displacement field ${\bf D}$ 
\begin{equation}\label{D}
  {\bf D}({\bf r}) =   
\varepsilon_0\widehat{\varepsilon}_h ({\bf r}) {\bf E}({\bf r}) + {\bf P}_{p}({\bf r})
+ {\bf P}_{\ell}({\bf r}) + {\bf P}_{m}({\bf r}).
\end{equation}
The electric field, then, can be written as
\begin{equation}\label{E}
{\bf E}({\bf r}) = \frac{ \widehat{\varepsilon}_h ^{-1}({\bf r})}{\varepsilon_0}({\bf D}({\bf r}) - {\bf P}_{p}({\bf r}) - {\bf P}_{\ell}({\bf r}) - {\bf P}_{m}({\bf r})).   
\end{equation}
For reducing the order of the second order differential equations for ${\bf P}_p$, ${\bf P}_{\ell}$, and ${\bf P}_m$ we introduce polarization currents
\begin{equation}\label{J}
    {\bf J}_n ({\bf r}) = \frac{\partial{\bf P}_n({\bf r})}{\partial t}, \; n = m, p, \ell.
\end{equation}
Combining Eqs.~\eqref{Four_level} with Eqs.~\eqref{Drude}-\eqref{E} and using Eq.~\eqref{J}, we obtain
following set of first-order differential and algebraic equations to be solved for modelling a four-level laser in photonic crystal -- liquid crystal -- metal microcavity
\begin{align}\label{Full_Set}
& \frac{\partial {{\bf J}}_{m}({\bf r})}{\partial t}+\kappa_{m}{\bf J}_{m}({\bf r}) = \varepsilon_0 \omega_{m}^2{\bf E}({\bf r}), \nonumber \\
& \frac{\partial{{\bf J}}_{p}({\bf r})}{\partial t}+\kappa_{p}{\bf J}_{p}({\bf r}) +
\bar{\omega}_{p}^2{\bf P}_{p}({\bf r})=-\frac{2\omega_{p}}{3\hbar}|{\bf d}_{p}|^2[\rho_3({\bf r})-\rho_0({\bf r})]{\bf E}({\bf r}), \nonumber \\
& \frac{\partial{{\bf J}}_{\ell}({\bf r})}{\partial t}+\kappa_{\ell}{\bf J}_{\ell}({\bf r})+
\bar{\omega}_{\ell}^2{\bf P}_{\ell}({\bf r})=-\frac{2\omega_{\ell}}{3\hbar}|{\bf d}_{l}|^2[\rho_2({\bf r})-\rho_1({\bf r})]{\bf E}({\bf r}), \nonumber \\
& \frac{\partial{\rho}_{3}({\bf r})}{\partial t}-\frac{1}{\hbar\omega_{p}}{\bf E}({\bf r})\cdot\left({\bf J}_{p}({\bf r}) + \frac{\kappa_{p}}{2}{\bf P}_{p}({\bf r})\right)
+
(\kappa_{0,3}+\kappa_{2,3}) \rho_3({\bf r})=
0, \nonumber \\
& \frac{\partial{\rho}_{2}({\bf r})}{\partial t}-\frac{1}{\hbar\omega_{\ell}}{\bf E}({\bf r})\cdot\left({\bf J}_{\ell}({\bf r})+\frac{\kappa_{\ell}}{2}{\bf P}_{\ell}({\bf r})\right)
+
\kappa_{1,2}\rho_2({\bf r})-\kappa_{2,3}\rho_3({\bf r})=0, \nonumber \\
& \frac{\partial{\rho}_{1}({\bf r})}{\partial t}+\frac{1}{\hbar\omega_{\ell}}{\bf E}({\bf r})\cdot\left({\bf J}_{\ell}({\bf r})+\frac{\kappa_{\ell}}{2}{\bf P}_{\ell}({\bf r})\right)
+
\kappa_{0,1} \rho_1({\bf r})-\kappa_{1,2}\rho_2({\bf r})=0, \nonumber \\
& \frac{\partial{\rho}_{0}({\bf r})}{\partial t}+\frac{1}{\hbar\omega_{p}}{\bf E}({\bf r})\cdot\left({\bf J}_{p}({\bf r})+\frac{\kappa_{p}}{2}{\bf P}_{p}({\bf r})\right)
-\kappa_{0,1}\rho_1({\bf r})-\kappa_{0,3}\rho_3({\bf r})=0, \nonumber \\
& \nabla\times{\bf E({\bf r})}=-\mu_0\frac{\partial {\bf H({\bf r})}}{\partial t}, \nonumber \\
& \nabla\times{\bf H}({\bf r})= \frac{\partial {\bf D}({\bf r})}{\partial t}, \nonumber \\
& {\bf E}({\bf r}) = \frac{ \widehat{\varepsilon}_h ^{-1}({\bf r})}{\varepsilon_0}({\bf D}({\bf r}) - {\bf P}_{p}({\bf r}) - {\bf P}_{\ell}({\bf r}) - {\bf P}_{m}({\bf r})), \nonumber \\
& {\bf J}_n ({\bf r}) = \frac{\partial{\bf P}_n({\bf r})}{\partial t}, \; n = m, p , \ell.
\end{align}

\newpage
\section{LINEAR RESPONSE}
\label{LINEAR RESPONSE}

Let us consider weak electromagnetic fields so that the non-linear terms in the third to sixth lines of Eq.~\eqref{Four_level} can be neglected. The time stationary solution, then, reads
\begin{align}
    & \rho_0({\bf r})=\rho({\bf r}), \ \rho_1({\bf r})=0, \ \rho_2({\bf r})=0, \rho_3({\bf r})=0.
\end{align}
After the time-harmonic substitution 
\begin{align}
{\bf E}(t) = \bm{E}(\omega)e^{-i\omega t}, \nonumber \\
{\bf P}_p(t)=\bm{P}_p(\omega) e^{-i\omega t}, \nonumber \\
{\bf P}_m(t)=\bm{P}_m(\omega) e^{-i\omega t},
\end{align}
the Eq.~\eqref{Drude} yields 
\begin{equation}
{\bm P}_m(\omega)=\varepsilon_0\chi_m\bm{E}(\omega),
\end{equation}
where Drude susceptibility reads as
\begin{equation}\label{Drude_Response}
\chi_m = -\frac{\omega_m^2}{\omega^2 + i\kappa_m\omega}.
\end{equation}
At the same time the first line in Eq.~\eqref{Four_level} yields 
\begin{equation}
{\bm P}_p(\omega)=\varepsilon_0\chi_p\bm{E}(\omega),
\end{equation}
where Lorentz susceptibility takes the following form
\begin{equation}\label{Lorentz_Response}
\chi_p=\frac{\omega_p}{\bar{\omega}_p^2-\omega^2-i\kappa_p\omega}
\frac{2|{\bf d}_p|^2\rho_0}{3\hbar\varepsilon_0}.
\end{equation}
Thus, we obtained the linear response solution in the form of the total dielectric permittivity  
\begin{equation}
\widehat{\varepsilon}_{\rm{tot}}=\widehat{\varepsilon}_h + \chi.
\end{equation}
The following parameters were used for Ag: $\omega_m = 13.7 \cdot 10^{15}$~rad$\cdot$s$^{-1}$, $\kappa_m = 27.35 \cdot 10^{12}$~rad$\cdot$s$^{-1}$, $\varepsilon_h = 5$~\cite{ordal1985optical}.
The following parameters were used for Pyrromethene 567:
$d_p = 6.5$~D, $d_{\ell} = 6.5$~D, $1 \text{D} = (1/299792458) \cdot 10^{-21}$~C$\cdot$~m, $\omega_p = 3.6 \cdot 10^{15}$~rad$\cdot$s$^{-1}$ ($\lambda_p = 524$~nm), $\omega_{\ell} = 3.4 \cdot 10^{15}$~rad$\cdot$s$^{-1}$ ($\lambda_{\ell} = 553$~nm), $\kappa_{0,3} = 5.14 \cdot 10^{9}$~rad$\cdot$s$^{-1}$, $\kappa_{2,3} = 1.07 \cdot 10^{14}$~rad$\cdot$s$^{-1}$, $\kappa_{1,2} = 5.14 \cdot 10^{9}$~rad$\cdot$s$^{-1}$, $\kappa_{0,1} = 1.62 \cdot 10^{14}$~rad$\cdot$s$^{-1}$,  $\rho = 8.17 \cdot 10^{24}$~m$^{-3}$.

\newpage
\section{FRANK-OSEEN MODEL}
\label{FRANK-OSEEN MODEL}

Let us consider a LC layer of thickness $L$. According to the Frank-Oseen model \cite{blinov2010structure} the LC free energy density can be written as
\begin{equation}\label{en_dens}
F=\frac{k_{11}}{2}(\nabla\cdot{\bf n})^2+\frac{k_{22}}{2}({\bf n}\cdot\nabla\times{\bf n}+2\pi/p_0)^2 +\frac{k_{33}}{2}({\bf n}\times \nabla\times{\bf n})^2 + w_{E},
\end{equation}
where $\bf n$ is the LC director, $k_{11}, k_{22}, k_{33}$ stand for splay, twist and bend elastic constants, respectively, and $p_0$ is the LC intrinsic pitch. 
The last term in Eq.~\eqref{en_dens} is the energy density of the quasi-static electric field
\begin{equation}\label{wE}
w_{E}=\frac{1}{8\pi}{\bf E}\cdot{\bf D}.
\end{equation}
The total free energy per unit surface of the LC cell is then
\begin{equation}\label{energy}
\frac{E}{S}=\int\limits_{z_1}^{z_2}Fdz+W, 
\end{equation}
where $W$ is the surface energy on the substrates. The director ${\bf n}$ can be written in the spherical coordinates as
\begin{equation}\label{dir}
{\bf n}=
\left(
\begin{array}{c}
\cos\theta\cos\phi \\
\cos\theta\sin\phi \\
\sin\theta
\end{array}
\right).
\end{equation}
Let us consider a nematic LC where $p_0 = \infty$.
On the substrates the director is fixed by rubbing in the $xy$-plane, so that $\phi(z) = 0$, see Fig.~\ref{Fig1:Model}.
Equating the variation of Eq.~\eqref{energy} to zero one obtains the Euler-Lagrange equations
\begin{align}\label{EL}
& \frac{d}{dz}\frac{\partial F}{\partial \theta'}-\frac{\partial F}{\partial \theta}=0.
\end{align}
The boundary conditions on the substrates are given by
\begin{align}\label{BC1}
& \left.\frac{\partial F}{\partial \theta'}\right|_{z=z_1, z_2}\mp\frac{\partial W}{\partial \theta_{1, 2}}=0, 
\end{align}
where
\begin{equation}
\theta(z_1):=\theta_1, \ \ \theta(z_2):=\theta_2.
\end{equation}
Now, our goal is to solve the above equations for unknown functions $\theta$ dependent on $z$.
First we write the dielectric tensor in terms of the LC director angle
\begin{equation}\label{LC_epsilon}
\hat{\epsilon} = \epsilon_{\scs{\perp}}\widehat{\mathbb{I}}+(\epsilon_{\scs{\parallel}}-\epsilon_{\scs{\perp}}){\bf n}\otimes{\bf n}.
\end{equation}
Since the displacement field ${\bf D}$ satisfies the Gauss theorem in the absence of free charges
one can write for 1D problem ${D}_z=\mathrm{const}$.
On the other hand, the electric vector ${\bf E}$ is the gradient of the electrostatic potential which can only be dependent on $z$, so that
$V=V(z)$ and ${\bf E} = E_z \hat{{\bm z}}$.
For the density of the electrostatic energy Eq.~\eqref{wE} we, thus, have
\begin{equation}\label{free}
w_{E}=\frac{1}{8\pi}\frac{D_z^2}
{\epsilon_{\scs{\perp}}\cos^2\theta+\epsilon_{\scs{\parallel}}\sin^2\theta}
\end{equation}
The increment of the electrostatic potential between the conducting layers
can be written as
\begin{equation}\label{BC_D}
V_2-V_2 = \int\limits_{z_1}^{z_2}dz\frac{D_z}{\epsilon_{\scs{\perp}}\cos^2\theta+\epsilon_{\scs{\parallel}}\sin^2\theta}.
\end{equation}
The latter equation is to be applied for specifying the unknown $D_z$.

The Euler-Lagrange equations, Eq.~\eqref{EL}, can be written in terms of the LC director as follows
\begin{align}\label{EL2}
& (k_{11}\cos^2\theta+k_{33}\sin^2\theta)~\theta''=(k_{11}-k_{33})\cos\theta\sin\theta(\theta')^2-
\frac{D_z^2}{4\pi}\frac{(\epsilon_{\scs{\parallel}}-\epsilon_{\scs{\perp}})\cos\theta\sin\theta}{(\epsilon_{\scs{\perp}}\cos^2\theta+\epsilon_{\scs{\parallel}}\sin^2\theta)^2}.
\end{align}
The boundary conditions on the substrates can be derived from the Rapini potential
\begin{equation}
W_{1,2}=\frac{W_0}{2}\sin^2(\theta_{1,2}-\theta_0),
\end{equation}
where $\theta_0$ is the pretilt angle.
Applying the Rapini potential in Eq.~\eqref{BC1} one writes
\begin{equation}\label{BC2}
(k_{11}\cos^2\theta_{1,2}+k_{33}\sin^2\theta_{1,2})\theta_{1,2}' \mp W_0\sin(\theta_{1,2}-\theta_0)\cos(\theta_{1,2}-\theta_0)=0.
\end{equation}

The non-dimentionalized, $z-z_0= tL$, differential equation can be written as 
\begin{align}\label{EL3}
&\theta'=\psi, \nonumber \\
&(\cos^2\theta+k_{3}\sin^2\theta)~\psi'=(1-k_{3})\cos\theta\sin\theta~\psi^2-
\frac{\mu^2}{4\pi}\frac{(\epsilon_{\scs{\parallel}}-\epsilon_{\scs{\perp}})\cos\theta\sin\theta}{(\epsilon_{\scs{\perp}}\cos^2\theta+\epsilon_{\scs{\parallel}}\sin^2\theta)^2}, 
\end{align}
where we defined
\begin{align}
 k_3=\frac{k_{33}}{k_{11}}, \
 \mu=\frac{D_zL}{\sqrt{k_{11}}}.
\end{align}
Together with the boundary conditions
\begin{align}
&[\cos^2\theta(0)+k_{3}\sin^2\theta(0)]\psi(0)=+\overline{W}\sin[\theta(0)-\theta_0]\cos[\theta(0)-\theta_0], \nonumber \\
&[\cos^2\theta(1)+k_{3}\sin^2\theta(1)]\psi(1)=-\overline{W}\sin[\theta(1)-\theta_0]\cos[\theta(1)-\theta_0],
\end{align}
where
\begin{equation}
\overline{W}=\frac{LW_0}{k_{11}},
\end{equation}
Eq.~\eqref{EL3} constitute a parametric boundary value problem for a set of two first order differential equations. The solution of this system has to be found consistently with the potential difference given by Eq.~\eqref{BC_D}, i.e.
\begin{align}\label{field}
U=\int\limits_{0}^{1}dt\frac{\mu}{\epsilon_{\scs{\perp}}\cos^2\theta+\epsilon_{\scs{\parallel}}\sin^2\theta},
\end{align}
where
\begin{equation}\label{U}
U=\frac{V_2-V_1}{\sqrt{k_{11}}}.
\end{equation}
As it can be found from Eq.~\eqref{field} the parameter $\mu$ has to be sampled as follows
\begin{equation}
\mu\in[\epsilon{\scs{\perp}}U,\epsilon_{\scs{\parallel}}U].
\end{equation}
In this work Eq.~\eqref{EL3} where solved by applying MATLAB R2014a in-built {\it bvb5c} solver (license No 984723). 
In every step the solver was initiated with $\mu=\epsilon{\scs{\parallel}}U$ calculated from a guess value of $U$. After the system was solved the true potential difference was computed from Eq.~\eqref{U}. The calculation was performed for the following 5CB parameters at a temperature of 25$^{\circ}$C:
$k11 = 5.9$~pN, $k33 = 9.9$~pN~\cite{nowinowski2012measurement}, $\varepsilon_{\scs{\perp}} = 6$ and $\varepsilon_{\scs{\parallel}} = 18$, $\theta_0 = 0.9^{\circ}$,
$W_0 = 0.00025$~Jm$^{-2}$.
For calculation the increment of the electrostatic potential between the conducting layers, the PVA layers with thickness 100 nm and $\varepsilon = 3$ were taken into account on both substrates.

\newpage
\section{FDTD}

The system of governing equations Eq.~\eqref{Full_Set} are solved using finite difference time domain method (FDTD)~\cite{taflove2005computational} according to Dr.~R.~C.~Rumpf's lectures~\cite{RumpfFDTD}.
All unknowns are defined on the Yee grid, staggered in space and time~\cite{yee1966numerical}, see Fig.~\ref{Fig:Yee} and Table~\ref{Tab:Yee}.
\begin{figure}[h!]
    \centering    \includegraphics[width=6cm]{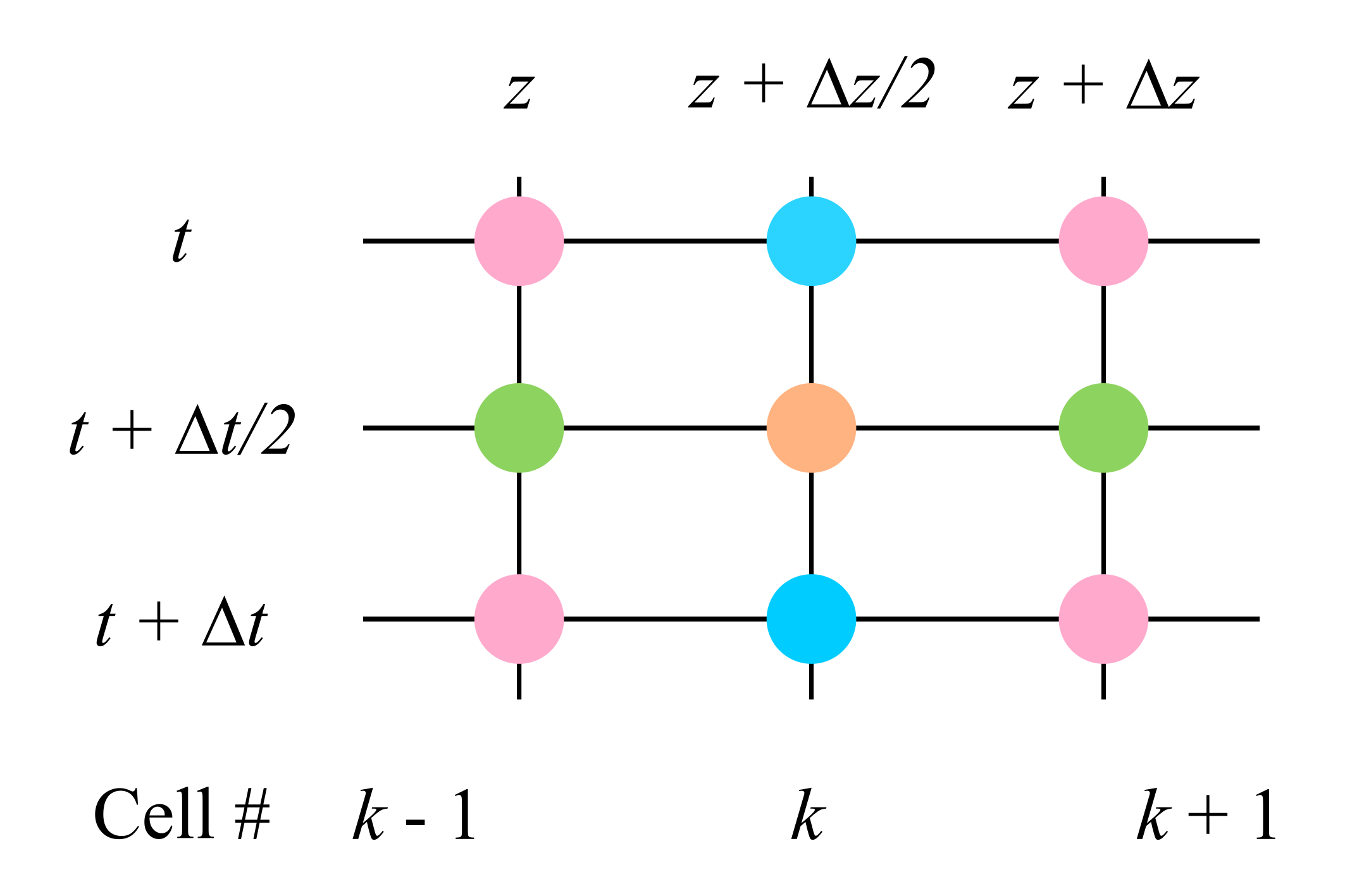}
    \includegraphics[width=4cm]{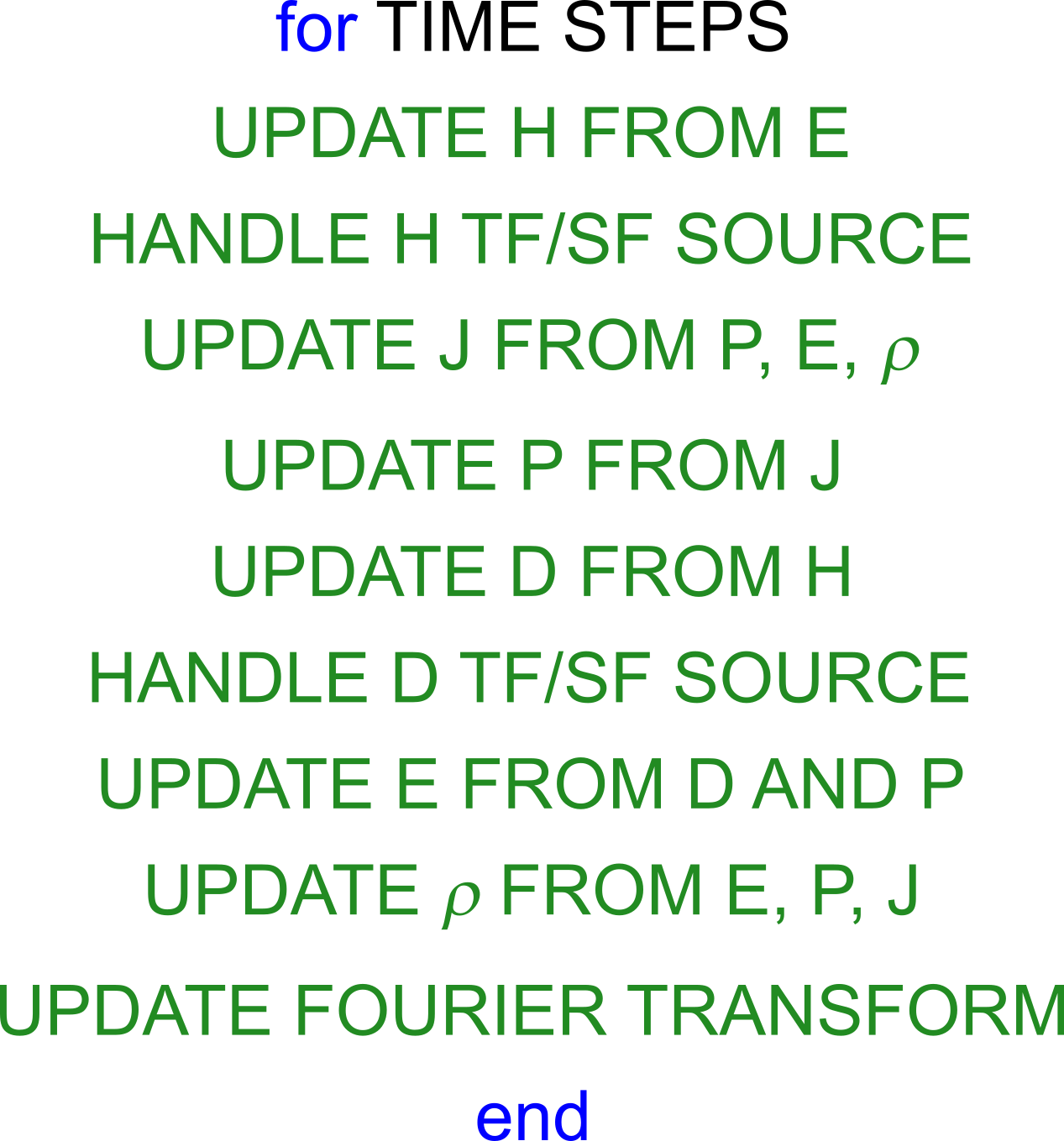} 
    \caption{1D Yee Grid (left)  FDTD loop sequence (right).}
    \label{Fig:Yee}
\end{figure}
\begin{table}[h!]
    \centering
    \begin{tabular}{|c|c|c|}
    \hline
     & $z$ & $z + \Delta z/2$ \\
     \hline
     $t$ & \cellcolor{1color} $E_x \; E_y \; D_x \; D_y  \; P_{m x} \; P_{p x} \; P_{\ell x} \; P_{m y} \; P_{p y} \; P_{\ell y} \; \rho_0 \; \rho_1 \; \rho_2 \; \rho_3$  &  \cellcolor{2color} $E_z \; P_{m z} \; P_{p z} \; P_{\ell z}$ \\
     \hline
     $t + \Delta t/2$ & \cellcolor{3color} $J_{m x} \; J_{p x} \; J_{\ell x} \; J_{m y} \; J_{p y} \; J_{\ell y}$ & \cellcolor{4color}  $H_x \; H_y \; J_{m z} \; J_{p z} \; J_{\ell z}$ \\
     \hline
    \end{tabular}
    \caption{Position of unknowns in time and space on Yee grid.}
    \label{Tab:Yee}
\end{table}
Let us treat the problem as one-dimensional, thus, all derivatives in Maxwell's curl equations with respect to $x$ and $y$ are taken as zero $\frac{\partial}{\partial x} = \frac{\partial}{\partial y} = 0$. Other derivatives are written as finite differences
\begin{equation}\label{Finite_Difference}
\frac{\partial f}{\partial z}\Biggr|_{z + \Delta z/2} = \frac{f(z + \Delta z) - f(z)}{\Delta z}; \quad \frac{\partial f}{\partial t}\Biggr|_{t + \Delta t/2} = \frac{f(t + \Delta t) - f(t)}{\Delta t},   
\end{equation}
while function values at the half-integer step are interpolated as arithmetic mean values
\begin{equation}\label{Interpolation}
f(z + \Delta z/2) = \frac{f(z + \Delta z) + f(z)}{2}; \quad f(t +\Delta t/2) = \frac{f(t + \Delta t) + f(t)}{2}.  
\end{equation}
Let us, for example, write the 7-th line of Eq.~\eqref{Full_Set} in the finite-difference form for $k$-th cell, using Eq.~\eqref{Finite_Difference}
\begin{equation}\label{FD_form}
\frac{\left.\rho_0\right|_{t+\Delta t} ^k-\left.\rho_0\right|_t ^k}{\Delta t}+\left.\frac{1}{\left.\hbar \omega_p\right|^k} {\bf E} {\bf J}_p\right|_{t+\Delta t/2} ^k + \left.\frac{\left.\kappa_p\right|^k}{\left.2 \hbar \omega_p\right|^k} {\bf E} {\bf P}_p\right|_{t+\Delta t/2} ^k-\left.\left.\kappa_{0,1}\right|^k \rho_1\right|_{t+\Delta t/2} ^k - \left.\left.\kappa_{0,3}\right|^k \rho_3\right|_{t+\Delta t/2}^k=0.
\end{equation}
Since all terms in Eq.~\eqref{FD_form} have to be taken at the same time step $t + \Delta t/2$ and coordinate $z$, the interpolation of unknowns has to be performed.
Using Eq.~\ref{Interpolation} and data from Fig.~\ref{Fig:Yee} and Table~\ref{Tab:Yee}, the dot product in the second term of Eq.~\eqref{FD_form} are written as
\begin{align}
\left.{\bf E} {\bf J}_p \right|_{t+\Delta t/2} ^k = \left.\left(E_x J_{p x}+E_y J_{p y}+E_z J_{p z}\right)\right|_{t+\Delta t/2}^k = 
\frac{\left.E_x\right|_{t+\Delta t} ^k+\left.E_x\right|_t ^k}{2} \left.J_{p x}\right|_{t + \Delta t/2} ^k + \nonumber \\
\frac{\left.E_y\right|_{t+\Delta t} ^k + \left.E_y\right|_t ^k}{2} \left.J_{p y}\right|_{t + \Delta t/2}^k +
\frac{\left.E_z\right|_{t+\Delta t}^k+\left.E_z\right|_t^k + \left.E_z\right|_{t+\Delta t}^{k-1}+ \left.E_z\right|_t^{k-1}}{4}\frac{\left.J_{p z}\right|_{t+\Delta t/2}^k + \left.J_{p z}\right|_{t+\Delta t/2}^{k-1}}{2}.
\end{align}
Expanding all terms in Eq.~\ref{FD_form} in the same way, the value $\left.\rho_0\right|_{t+\Delta t} ^k$, finally, can be expressed in terms of other unknowns. In the same manner, all unknowns in Eq.~\ref{Full_Set} at the future time step are expressed in terms of unknowns at previous time steps at all number of cells $N$.

To prevent reflection from the numerical domain boundaries, the perfect boundary condition is applied.
Time step $\Delta t$ is chosen so physical waves travel one cell near the boundary in exactly two time steps
$\Delta t = \sqrt{\varepsilon_b}\Delta z/2c$, where $\varepsilon_b$ is the permittivity at both boundaries and $c$ is the speed of light.
That choice of time step also allows us to satisfy the Courant stability condition.
To implement the perfect boundary condition, the curl Maxwell's equations at the 1-st and $N$-th cells are written as
\begin{align}
\left.H_x\right|_{t+\Delta t/2}^N = \left.H_x\right|_{t - \Delta t/2}^N + \frac{c \Delta t}{\Delta z}\left( \left.E_y\right|_{t - 2\Delta t}^N - \left.E_y\right|_{t}^N\right) \nonumber \\
\left.H_y\right|_{t+\Delta t/2}^N = \left.H_y\right|_{t - \Delta t/2}^N - \frac{c \Delta t}{\Delta z}\left( \left.E_x\right|_{t - 2\Delta t}^N - \left.E_x\right|_{t}^N\right) \nonumber \\
\left.D_x\right|_{t+\Delta t}^1 = \left.D_x\right|_{t}^1 - \frac{c \Delta t}{\Delta z}\left( \left.H_y\right|_{t + \Delta t/2}^1 - \left.H_y\right|_{t - 3\Delta t/2}^1\right) \nonumber \\
\left.D_y\right|_{t+\Delta t}^1 = \left.D_y\right|_{t}^1 + \frac{c \Delta t}{\Delta z}\left( \left.H_x\right|_{t + \Delta t/2}^1 - \left.H_x\right|_{t - 3\Delta t/2}^1\right).
\end{align}
The total field (TF)/scattered field (SF) technique is used for incorporating EM field source.
The Gaussian pulse source is used for calculating spectra in the case of linear response regime.
The sinusoidal wave with the frequency $\omega$ is used as source for pumping.
To satisfy causality, the FDTD loop has to be performed in a certain sequence, see Fig.~\ref{Fig:Yee}.
To receive spectra in linear response regime and PL intensity spectra, the Fourier transform $\mathcal{F}(E)$ is calculated at 1-st cell (SF region) for reflectance and $N$-th cell (TF region) for transmittance.
In the linear response regime the reflectance is defined as follows $\text{R} = \mathcal{F}(E)/\mathcal{F}(E_{src})$, where $E_{src}$ is the source field.
The simulation time was equal to $4/\Delta f$, where $\Delta f$ is the linewidth of the resonant line with the highest Q-factor. 
The FDTD algorithm was implemented in MATLAB R2014a (license No 984723).


\end{widetext}

\bibliography{main.bib}

\end{document}